\documentclass[useAMS,fleqn,usenatbib]{mnras}
\usepackage[utf8]{inputenc}
\usepackage{amssymb,amsmath}
\usepackage{graphicx}
\usepackage[dvipsnames]{xcolor}
\usepackage[none]{hyphenat}

\usepackage{newtxtext,newtxmath}

\usepackage{array}
\newcolumntype{P}[1]{>{\centering\arraybackslash}p{#1}}

\newcommand{\de}{\mathrm{d}}

\title[Ruling out Strongly Interacting DM–DR Models]{Ruling out Strongly Interacting Dark Matter–Dark Radiation Models from Joint Observations of Cosmic Microwave Background and Quasar Absorption Spectra}

\author[Chatterjee, Mitra, Banerjee]{
Atrideb Chatterjee$^{1}$,
Sourav Mitra$^{2}$\thanks{E-mail:hisourav@gmail.com},
Amrita Banerjee$^{3}$
\\
$^{1}$ Inter-University Centre for Astronomy and Astrophysics, Post Bag 4, Ganeshkhind, Pune 411007, India\\
$^{2}$ Department of Physics, Surendranath College, 24/2 M. G. Road, Kolkata 700009, India\\
$^{3}$ Department of Physics, Visva-Bharati University, Santiniketan, 731235, West Bengal, India\\
}

\date{Accepted XXX. Received YYY; in original form ZZZ}

\pubyear{2023}

\begin{document}
\def\cyan{\color{cyan}}
\def\red{\color{red}}
\def\blu{\color{blue}}
\def\refjnl#1{{\rm#1}}
\def\aj{\refjnl{AJ}}                   
\def\actaa{\refjnl{Acta Astron.}}      
\def\araa{\refjnl{ARA\&A}}             
\def\apj{\refjnl{ApJ}}                 
\def\apjl{\refjnl{ApJ}}                
\def\apjs{\refjnl{ApJS}}               
\def\ao{\refjnl{Appl.~Opt.}}           
\def\apss{\refjnl{Ap\&SS}}             
\def\aap{\refjnl{A\&A}}                
\def\aapr{\refjnl{A\&A~Rev.}}          
\def\aaps{\refjnl{A\&AS}}              
\def\azh{\refjnl{AZh}}                 
\def\baas{\refjnl{BAAS}}               
\def\bac{\refjnl{Bull. astr. Inst. Czechosl.}} 
\def\caa{\refjnl{Chinese Astron. Astrophys.}} 
\def\cjaa{\refjnl{Chinese J. Astron. Astrophys.}} 
\def\icarus{\refjnl{Icarus}}           
\def\jcap{\refjnl{J. Cosmology Astropart. Phys.}} 
\def\jrasc{\refjnl{JRASC}}             
\def\memras{\refjnl{MmRAS}}            
\def\mnras{\refjnl{MNRAS}}             
\def\na{\refjnl{New A}}                
\def\nar{\refjnl{New A Rev.}}          
\def\pra{\refjnl{Phys.~Rev.~A}}        
\def\prb{\refjnl{Phys.~Rev.~B}}        
\def\prd{\refjnl{Phys.~Rev.~D}}        
\def\pre{\refjnl{Phys.~Rev.~E}}        
\def\prl{\refjnl{Phys.~Rev.~Lett.}}    
\def\pasa{\refjnl{PASA}}               
\def\pasp{\refjnl{PASP}}               
\def\pasj{\refjnl{PASJ}}               
\def\rmxaa{\refjnl{Rev. Mexicana Astron. Astrofis.}} 
\def\qjras{\refjnl{QJRAS}}             
\def\skytel{\refjnl{S\&T}}             
\def\solphys{\refjnl{Sol.~Phys.}}      
\def\sovast{\refjnl{Soviet~Ast.}}      
\def\ssr{\refjnl{Space~Sci.~Rev.}}     
\def\zap{\refjnl{ZAp}}                 
\def\nat{\refjnl{Nature}}              
\def\iaucirc{\refjnl{IAU~Circ.}}       
\def\aplett{\refjnl{Astrophys.~Lett.}} 
\def\apspr{\refjnl{Astrophys.~Space~Phys.~Res.}} 
\def\bain{\refjnl{Bull.~Astron.~Inst.~Netherlands}}  
\def\fcp{\refjnl{Fund.~Cosmic~Phys.}}  
\def\gca{\refjnl{Geochim.~Cosmochim.~Acta}}   
\def\grl{\refjnl{Geophys.~Res.~Lett.}} 
\def\jcp{\refjnl{J.~Chem.~Phys.}}      
\def\jgr{\refjnl{J.~Geophys.~Res.}}    
\def\jqsrt{\refjnl{J.~Quant.~Spec.~Radiat.~Transf.}} 
\def\memsai{\refjnl{Mem.~Soc.~Astron.~Italiana}} 
\def\nphysa{\refjnl{Nucl.~Phys.~A}}   
\def\physrep{\refjnl{Phys.~Rep.}}   
\def\physscr{\refjnl{Phys.~Scr}}   
\def\planss{\refjnl{Planet.~Space~Sci.}}   
\def\procspie{\refjnl{Proc.~SPIE}}   
\let\astap=\aap
\let\apjlett=\apjl
\let\apjsupp=\apjs
\let\applopt=\ao

\def\der{{\rm d}}
\label{firstpage}
\pagerange{\pageref{firstpage}--\pageref{lastpage}}
\maketitle

\begin{abstract}
The cold dark matter (CDM) paradigm provides a remarkably good description of the Universe's large-scale structure. However, some discrepancies exist between its predictions and observations at very small sub-galactic scales. To address these issues, the consideration of a strong interaction between dark matter particles and dark radiation emerges as an intriguing alternative. In this study, we explore the constraints on those models using joint observations of Cosmic Microwave Background (CMB) and Quasars absorption spectra with our previously built parameter estimation package \texttt{CosmoReionMC}. At 2-$\sigma$ confidence limits, this analysis rules out the strongly interacting Dark Matter - Dark Radiation models within the recently proposed ETHOS framework, representing the most stringent constraint on those models to the best of our knowledge. Future research using a 21-cm experiment holds the potential to reveal stronger constraints or uncover hidden interactions within the dark sector. 

\end{abstract}


\section{Introduction}

Dark Matter (DM) remains an enigma even today despite being one of the main constituents of the structure formation of the Universe. While the widely accepted $\Lambda$CDM (Cold Dark Matter) model is extremely successful in explaining large scale structures, it has several shortcomings in small-scales, such as {\it (i)} observations related to the rotational curves of dwarf galaxies with similar stellar masses $ (M_{*} \geq 10^7 M _{\odot}) $ show a variety of inner structure of the dark matter profile which are in strong disagreement with the N-body simulation \citep{2015MNRAS.452.3650O}  {\it (ii)} Too big to fail problem for the field galaxies \citep{boylan2011, boylan2012, 2016MNRAS.460.3610O} {\it (iii)} Observations of dwarf galaxies with a fixed stellar mass show a variety of sizes  which is yet unexplained within the $\Lambda$CDM paradigm \citep{2022NatAs...6..897S}  {\it (iv)} Observations from the SAGA survey \citep{2017ApJ...847....4G, 2021ApJ...907...85M} point towards a considerably lower fractions of dormant satellites and are in stark contrast with state of the art cosmological zoom-in simulations at $M_{*} \leq 10^8 M_{\odot}$ \citep{2019A&A...624A..11H, 2022NatAs...6..897S}. There is a possibility that some of these problems can be resolved with more observational data and/or a more self-consistent treatment of baryonic feedback or a higher-resolution hydrodynamic simulation. However, these observations are in strong tension with the state-of-the-art simulations.
A viable solution to this problem is to invoke additional non-CDM models. These models can eliminate the small-scale power inherent in the $\Lambda$CDM model and solve the CDM model's small-scale issues without altering the more complicated baryonic physics. \citep{PhysRevLett.72.17, 2012MNRAS.420.2318L}

To this end, a novel framework called the effective theory of structure formation (ETHOS) has been introduced, which forms a bridge joining a broad range of DM particle physics with the structure formation in the Universe \citep{2016PhRvD..93l3527C}.
One such type of DM model where the DM is coupled to the relativistic species (commonly known as the ``dark radiation") is called Dark Acoustic Oscillations (DAO) model. Later, a modified framework based on ETHOS has been proposed \citep{2020MNRAS.498.3403B, 2021MNRAS.506..128B} where two effective parameters $h_{\rm peak}$ (the first DAO peak's amplitude) and $k_{\rm peak}$ (the wave number associated with the first DAO peak) controls the main features of the model. In this set-up, $h_{\rm peak}\rightarrow 0$ represents a Warm Dark Matter (WDM) model, whereas $k_{\rm peak}\rightarrow \infty$ indicates CDM. 

A number of works have been put forward in recent times \citep{2023arXiv230406742S, 2022MNRAS.516.1524K, 2021PhRvD.103d3512M, 2021MNRAS.506..128B, 2021MNRAS.504.3773S, 2019ApJ...874..101S, 2018MNRAS.477.2886L, 2016MNRAS.460.1399V} to explore the effect of DAO model on high-redshift observations. In particular, using state-of-the-art hydrodynamical simulations \cite{2022MNRAS.516.1524K} has tried to quantify if there exist any observable differences between $\Lambda$CDM and these DAO models \footnote{note that they also incorporate the DM-DM interaction in their model} from the upcoming JWST observations, but their results remain inconclusive. Further, \cite{2019JCAP...10..055A, PhysRevD.98.083540} have investigated in detail the viability of the DAO models as an alternative to solve the small-scale issue and carried out a joint analysis to put constraints on these models with Cosmic Microwave Background (CMB), Baryonic Acoustic Oscillation (BAO), and Lyman-$\alpha$ data. Their results when translated into $h_{\rm peak} - k_{\rm peak}$ framework, indicate that the DAO with $h_{\rm peak}$ in the range of $0.2-0.6$ and $k_{\rm peak}$ of $40-80~h {\rm Mpc^{-1}}$  are ruled out \citep{2020MNRAS.498.3403B}.
Interestingly, the {\it strongly} coupled DAO models, known as sDAO, with $h_{\rm peak}$ in the range $0.8-1.0$, are still allowed and non-degenerate with CDM or WDM up to $z=5$ for $k_{\rm peak}<60~h {\rm Mpc^{-1}}$. Moreover, this parametrisation can accurately describe the linear power spectrum and the halo mass function up to $z \geq 5$ for such sDAO models.

However, the compatibility of these models with high-redshift observations related to reionization is yet to be tested. In fact, Ultraviolet (UV) photons from distant stars and Quasars (dominant source for helium reionization) ionize the neutral intergalactic medium (IGM), which can, in principle, put constraints on those DAO models. This is due to the fact that the suppression of low-mass dark matter halos in these models would result in fewer ionizing photons, thereby delaying the history of reionization \citep{2018JCAP...08..045D}. To this aim, this {\it lette}r focuses on checking the viability of the wide range of sDAO models, which are otherwise allowed, by employing an MCMC-based reionization model \texttt{CosmoReionMC} \citep{2021MNRAS.507.2405C} (referred to as CCM21 hereafter) utilizing observations of both CMB and Quasar absorption spectra while simultaneously varying all cosmological and astrophysical parameters.

\section{Theoretical Model}
\subsection{halo mass function for sDAO}
\label{sec:nCDM_transfer} 

We will start by deriving the halo mass function for the sDAO model within the ETHOS framework. Based on the Extended Press-Schechter formalism, the halo mass function can be calculated as \citep{1999MNRAS.308..119S}:
\begin{equation}
    \frac{\partial n(M, z)}{\partial M}=  \frac{\Bar{\rho}_{m}}{M^2} \left|\frac{\der \log \sigma}{\der\log M}\right| f(\nu) 
\end{equation} 
where $\Bar{\rho}_{m}$ is the background matter density
of the Universe at $z=0$,  $\sigma$ is the linear mass variance, $\nu \equiv  \frac{1.686^2}{\sigma^2(R)D^2(z)} $
with $D(z)$ being the growth function and R is the halo
radius enclosing the mass M. Further, $f(\nu)$ is given by
\begin{equation}
    f(\nu)=A\sqrt{\frac{2q\nu}{\pi}} \left[ 1+ \left(q \nu \right)^{-p} \right]\exp \left[ -\frac{q\nu}{2} \right]
\end{equation}
with $A = 0.3658$, $q = 1.0$, $p = 0.3$ obtained from \cite{2021MNRAS.506..128B}.

Here the variance is given by
\begin{equation}
    \sigma^2(R) = \frac{1}{2 \pi^2} \int_{0}^{\infty} dk k^2 P_{\rm sDAO}(k) \tilde{W^2}_{R} (k)
\end{equation}
where $\tilde{W}_{R} (k)$ is Fourier transform of the window function given by \citep{2019ApJ...874..101S}
\begin{equation}
    \tilde{W}_{R} (k) = \frac{1}{1+ \left( \frac{kR}{c_{w}} \right)^{\delta}}
\end{equation}
where $\delta = 3.46$ and $c_{w} = 3.79$ \citep{2021MNRAS.506..128B}. On the other hand, $P_{\rm sDAO}$, denoting the power spectrum of the strongly interacting DM-DR model, is calculated as \citep{2020MNRAS.498.3403B}
\begin{equation}
    P_{\rm sDAO} (k) = T^{2}_{L} (k) P_{\rm CDM}(k)
\end{equation}
where $T_{L}$ is the linear transfer function whose form mostly depends on the location $k_{\rm peak}$ and amplitude $h_{\rm peak}$ of the first DAO peak. In this letter, we have carried out an analysis for sDAO with $h_{\rm peak} = 1.0$ and $h_{\rm peak} = 0.8$ treating $k_{\rm peak}$ as a free parameter. The other fitting parameters for the transfer function have been obtained from Table-2 of \cite{2020MNRAS.498.3403B}.

\subsection{Modelling Reionization}

In this work, we apply the semi-analytical data-constrained reionization model CCM21 which is based on \cite{2005MNRAS.361..577C,2006MNRAS.371L..55C}. In brief, we will summarise here this model's essential features in what follows. 

In this model, a set of coupled ordinary differential equations is solved to compute both Hydrogen and Helium's ionization and thermal histories simultaneously and consistently. The inhomogeneities of the IGM are modelled following the analytical calculation presented in \cite{2000ApJ...530....1M} where the reionization is taken to be completed once all the low-density regions with overdensities $\Delta_i<\Delta_{\rm c}$ are ionized, where $\Delta_{\rm c}$ is the critical density. Here, it is sufficient to take Quasars and population II (PopII) stars as the ionizing sources \citep{2021MNRAS.507.2405C}. We calculate the Quasar contribution from the observed Quasar luminosity function (LF) at $z<7.5$ \citep{2019MNRAS.488.1035K}. Note that, Quasars do not dominate the photon budget at higher redshifts and thus their contribution to hydrogen reionization is not significant. The stellar (PopII) contribution is calculated as

\begin{equation}
\dot{n}_{\rm ph, \rm PopII}(z) 
      =\mathcal{E} \rho_b  \frac{\der f_{\mathrm{coll}}}{\der t} \int^{\infty}_{\nu_H}  \left(\frac{\der N_{\nu}}{\der M} \right) \der \nu
\end{equation}

where $\rho_b$ is the mean comoving density of baryons in the IGM, $\nu_{H}$ is the threshold frequency for hydrogen photoionization, and $\mathcal{E} = f_{*} \times f_{\mathrm{esc}}$, where $f_*$ is the star formation efficiency, $f_{\mathrm{esc}}$ is the escape fraction of the ionizing photons. The quantity $\der N_{\nu}/ \der M$ denotes the number of photons emitted per frequency range per unit mass of the star and depends on the stellar spectra and initial mass function (IMF) of the stars \citep{2005MNRAS.361..577C}.
Applying a Saltpeter IMF in the mass range $1-100 M_{\odot}$ with a metallicity of $0.05 M_{\odot}$, $\der N_{\nu}/ \der M$ has been obtained from the stellar synthesis model of \cite{2003MNRAS.344.1000B}.
In principle, both $f_*$ and $f_{\rm esc}$ may depend on redshifts and halo masses \citep{2013MNRAS.428L...1M, 2015MNRAS.454L..76M, 2017MNRAS.472.2009Q, 2018MNRAS.479.4566M,2020MNRAS.491.3891P,2023MNRAS.523L..35M}. However, we decide to ignore their mass and redshift dependencies in the present analysis to keep our model simple. Finally, $\der f_{\mathrm{coll}}/\der t$ is the rate of collapse fraction of the dark matter halos and is calculated as \citep{2005MNRAS.361..577C}:
\begin{equation}
        f_{\mathrm{coll}} = \frac{1}{\Bar{\rho}_{m}}\int^{\infty}_{M_{\mathrm{min}}(z)} \de M  M \frac{\partial n(M,z)}{\partial M},
\end{equation}
where $M_{\rm min}(z)$ is the minimum mass for star-forming halos.

Furthermore, the mean free path of ionizing photons is computed using an analytical prescription given by \citep{Miralda00} where reionization is complete when all the low-density regions have been ionized. The computation of the mean free path involves a normalization parameter $\lambda_{0}$, considered as a free parameter of the model and later constrained using the observations of the redshift distribution of the Lyman Limit systems \citep{2005MNRAS.361..577C}.

\section{Datasets \& Likelihood}
In order to put joint constraints on the sDAO models using observations of both CMB and Quasar, we perform an MCMC analysis by varying all the cosmological and astrophysical parameters simultaneously (referred to as \textbf{CMB+Quasar} hereafter). 
For this purpose, we have modified the \texttt{CosmoReionMC} package described in CCM21 in order to incorporate the effect of sDAO based on the non-CDM framework of \citep{2020MNRAS.498.3403B, 2021MNRAS.506..128B}. The logarithm of total likelihood function is given as
\begin{equation}
    \mathcal{L}=\mathcal{L}_{\mathrm{Pl}} + \mathcal{L}_{\mathrm{Re}},
\end{equation}
where $\mathcal{L}_{\mathrm{Pl}}$ is the log-likelihood function corresponding to the Planck 2020 observations \citep{2020A&A...641A...6P} and
\begin{equation}
\mathcal{L}_{Re}=\frac{1}{2}\sum^{N_{\mathrm{obs}}}_{\alpha=1}\left[\frac{\zeta^{\mathrm{obs}}_{\alpha}-\zeta^{\mathrm{th}}_{\alpha}}{\sigma_{\alpha}}\right]^2.
\end{equation}
Here $\zeta^{\mathrm{obs}}_{\alpha}$ is the set of $N_{\mathrm{obs}}$ observational data related to Quasars which are as follows. $(i)$ Combined analysis of Quasar absorption spectra and hydrodynamic simulations of photoionization rates $\Gamma_{\mathrm{PI}}$ \citep{2013MNRAS.436.1023B, 2018MNRAS.473..560D}, $(ii)$ the redshift distribution of Lyman-limit system $\de N_{\mathrm{LL}} / \de z$ \citep{2010ApJ...718..392P, 2013ApJ...765..137O, 2019MNRAS.482.1456C}, and $(iii)$  a measurement of the upper limit on the neutral hydrogen fractions obtained from the dark fractions in Quasar spectra \citep{2015MNRAS.447..499M}. The later data has been used as a prior while calculating the likelihood along with the condition that reionization has to be completed ($Q_{\mathrm{HII}} = 1$) by $z \sim 5.3$ \citep{2015MNRAS.447.3402B, 2018MNRAS.479.1055B, 2018ApJ...864...53E, 2021MNRAS.501.5782C}. The $\sigma_{\alpha}$ represents the error bars coming from these observations, and the $\zeta^{\mathrm{th}}_{\alpha}$ are the theoretical predictions obtained from our model.

For the CMB anisotropy calculation, we follow the exact same procedure as outlined in CCM21. We modify the publicly available Boltzmann solver code CAMB \citep{Lewis:2013hha} so that the reionization history from our model can be incorporated, replacing the default reionization set-up of the CAMB. 

The free parameters for this analysis are
\begin{equation}
    \Theta =\{H_{0}, \Omega_{b}h^{2}, \Omega_ch^{2}, A_{s}, n_s, \mathcal{E}, \lambda_{0}, k^{-1}_{\rm peak} \},
\end{equation}
While $H_{0}, \Omega_{b}h^{2}, \Omega_ch^{2}, A_{s}, n_s$ constitutes the set of cosmological parameters, $\mathcal{E}, \lambda_{0}$ are the  astrophysical parameters appeared in our model. Finally, $k^{-1}_{\rm peak}$ denotes the inverse of the location of the first peak of the DAO model in the unit of $h^{-1} {\rm Mpc}$.

For the MCMC run, we take a broad flat prior for all eight free parameters to avoid possible biases arising due to the initial choices of parameter values. For $k^{-1}_{\rm peak}$, the prior range is taken as $[0.001, 1.0 ]$ allowing us to explore $k_{\rm peak}$ in the range $1-1000$ $h {\rm Mpc^{-1}}$. For sampling the parameter space with MCMC run, we use 32 walkers and continue the run in parallel mode using 16 cores until the convergence criteria (as described in CCM21 and \citealt{2013PASP..125..306F}) is met.

\section{Results}
\label{sec:result}
\begin{figure}
    \centering
    \includegraphics[width =\columnwidth]{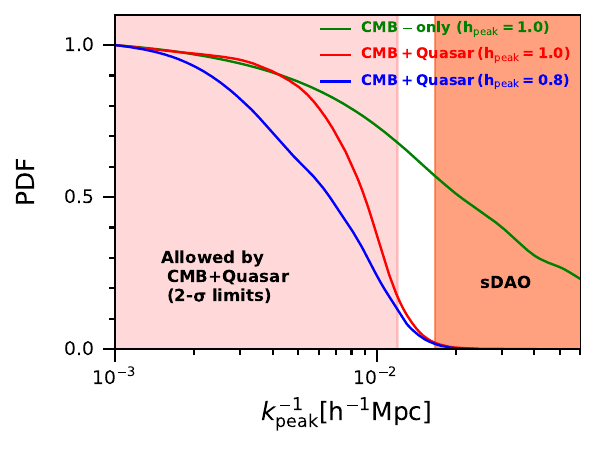}
    \caption{1D marginalized posterior distribution of the constraints on $k^{-1}_{\rm peak}$. The green curve represents the case when only CMB observations are used. The red and blue curve corresponds to the case for \textbf{CMB+Quasar} case for $h_{\rm peak} = 1.0$ and $h_{\rm peak} = 0.8$ respectively. The orange-shaded region in the {\it right side} shows the range of $k^{-1}_{\rm peak}$ values where all the sDAO model lies, whereas the salmon-shaded region in {\it left} shows the range of $k^{-1}_{\rm peak}$ allowed by the \textbf{CMB+Quasar} case up to the 2-$\sigma$ level. This is same for both $h_{\rm peak} = 1.0$ and $h_{\rm peak} = 0.8$.}
    \label{fig:k_peak_inv}
\end{figure}

\begin{figure*}
    \centering
    \includegraphics[width = \textwidth]{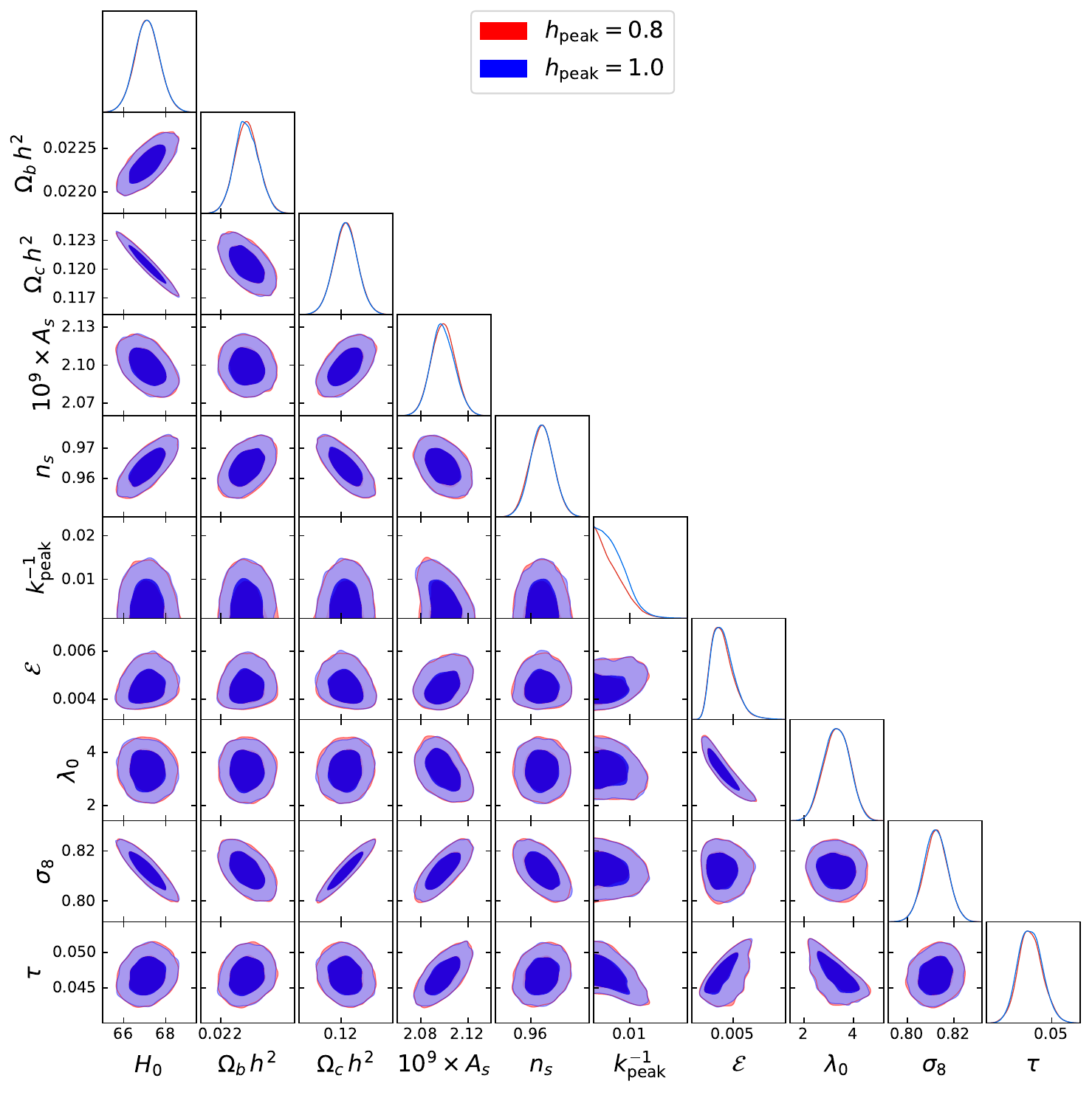}
    \caption{The marginalized posterior distribution of  all the free parameters and 2 derived parameters (CMB optical depth $\tau$ and the variance at $8 h^{-1} {\rm Mpc}$ i.e. $\sigma_8$) obtained in \textbf{CMB+Quasar} case. The red contour shows the scenario for $h_{\rm peak} = 1.0$, whereas the blue contour depicts the $h_{\rm peak} = 0.8$ case. Two-dimensional plots in the figure show the joint probability distribution (1-$\sigma$ and 2-$\sigma$ contours) of any two free parameters. It is evident that $k^{-1}_{\rm peak}$ shows no correlation with either astrophysical parameters or cosmological parameters.}
    \label{fig:diff_hpeak}
\end{figure*}

The 1D marginalized posterior distribution of the $k^{-1}_{\rm peak}$ for  $h_{\rm peak} = 1.0$ and $h_{\rm peak} = 0.8$ are shown (in red and blue respectively) in Figure \ref{fig:k_peak_inv}. The figure shows that the \textbf{CMB+Quasar} analysis allows sDAO models only with $k_{\rm peak} > 83$ $h {\rm Mpc^{-1}}$ at $95\%$ confidence level and clearly rules out others with smaller $k_{\rm peak}$ values. To make it visually more clear, the orange-shaded regions in the figure represent the area where the sDAO model would lie in the \textbf{CMB+Quasar} analysis. This result is the most stringent constraint on the sDAO model to the best of our knowledge. As the constraints on $k_{\rm peak}$ are the same for both $h_{\rm peak} = 1.0$ and $h_{\rm peak} = 0.8$, this result implies that any sDAO models with $h_{\rm peak}$ in the range between $0.8 - 1.0$ can also be ruled out. This essentially means that all the sDAO models inside the brown shaded region in the upper right corner of the Figure 10 of \citep{2020MNRAS.498.3403B} are effectively ruled out when observations from Quasars are used.

In order to distinguish the effect of CMB data from the Quasars, we carried out a separate analysis with only-CMB observations. The constraints on $k^{-1}_{\rm peak}$ from this \textbf{only-CMB} analysis has been shown in Figure \ref{fig:k_peak_inv} as a green line. In this case, all sDAO model with $k_{\rm peak} > 16$ $h {\rm Mpc^{-1}}$ are allowed at $95\%$ confidence. In comparison to \textbf{CMB+Quasar} scenario, the constraints from CMB alone are considerably weaker. It is immediately evident that the Quasar observations have played a crucial role in strengthening the constraints placed on the sDAO models.

For a detailed understanding of the constraints on different parameters and their correlation for the \textbf{CMB+Quasar} scenario, the posterior distribution of those is shown in Figure \ref{fig:diff_hpeak}. We should mention that the CMB optical depth $\tau$ and the variance at $8$ $h^{-1} {\rm Mpc}$ $\sigma_8$ presented in the figure are the derived parameters of this analysis. It is evident from the figure that $k_{\rm peak}$ is not strongly correlated with any of the other free parameters involved in this analysis, as any variation in $k_{\rm peak}$ can always be compensated for by variations in reionization related parameters, e.g. $\mathcal{E} $ and $\lambda_{0}$.

\section{Conclusions}

A number of particle physics models predict Dark Matter (DM)-Dark Radiation (DR) interactions, including a variety of dark matter species and novel forces as a viable solution to the {\it small-scale problems} present in the current $\Lambda$CDM model. A hallmark of such interactions is embedded as oscillatory features in the linear matter power spectrum, which are known as Dark Acoustic Oscillations (DAOs). The ETHOS framework is one of the powerful tools to couple dark matter physics with the structure formation of the Universe. Especially, the two-parameter model with the location $k_{\rm peak}$ and amplitude $h_{\rm peak}$ of the first DAO peak in the ETHOS paradigm provides an unprecedented approach to capture the main features of DM-DR interactions. However, it remains to be seen whether these models are compatible with observations from Quasars at high redshifts, which can have direct consequences on the ionization history of the Universe occurring somewhere between redshifts $z\sim6-15$. 

It is important to mention some of the caveats of this analysis. Here we assume a fixed non-evolving value for the parameter $\mathcal{E}$, which could potentially oversimplify the reionization scenario. This oversimplification may arise from the possibility that this parameter could be dependent on halo mass and/or redshift, leading to a weaker constraint on the allowed sDAO models. The delay in structure formation in sDAO models may have been compensated if the $\mathcal{E}$ were allowed to vary with redshift and halo properties. In fact, previous studies \citep{2017ApJ...836...16D, 2017PhRvD..96j3539L} have revealed that reionization and structure formation may initiate later in non-CDM models but eventually catch up to CDM models in their final stages due to the faster baryonic assembly. However, introducing $\mathcal{E}$ as a function of redshift and halo mass would lead to an additional larger number of free parameters (e.g. see \citealt{2023MNRAS.523L..35M}) in our model, thus making the MCMC simulation challenging to converge.
Secondly, accurate modelling of small-scale morphology of reionization processes might be needed for the robust prediction of faint low-mass galaxies' abundance \citep{2023MNRAS.525.5932B, 2023arXiv230406742S} which can potentially affect the constraints on the sDAO models. However, this detailed morphological study is beyond the scope of this semi-analytical model. Furthermore, the parameter $\mathcal{E}$ might have mild dependencies on a number of factors including the initial mass function (IMF), stellar age and metallicity, and the dust content. These effects have been ignored in order to keep the analysis simple.

For further investigation and constraint of such models, we have studied here the impact of joint CMB and high-redshift Quasar observations on the sDAO models using a data-constrained reionization scenario. We have coupled our previously developed \texttt{CosmoReionMC} package with the sDAO models in the ETHOS framework and obtained the most stringent constraint on them. We find that the constraints become significantly weak when only CMB observations are used, whereas, with the addition of Quasar data, all the sDAO models proposed within the ETHOS framework are ruled out at 2-$\sigma$ limits. Future research using a 21-cm experiment may reveal stronger constraints or uncover hidden dark sector interactions.

\section*{Acknowledgements}
We thank the anonymous reviewer for providing valuable and constructive feedback on the manuscript.
AC thanks Pratika Dayal for insightful comments on the draft. AC also wishes to acknowledge the computing facility provided by IUCAA.

\section*{Data Availability}
The observational datasets used here are taken from the literature, and the code used for this work will be shared on request with the corresponding author.

\label{lastpage}

\bibliographystyle{mnras}
\bibliography{reion}

\section*{Supporting Information}
Supplementary materials are available at \href{https://academic.oup.com/mnrasl/article/528/1/L168/7478034#supplementary-data}{MNRASL} online.

\end{document}